\documentclass[3p,twocolumn]{elsarticle}
\usepackage{lineno,hyperref}

\journal{Information Fusion}

\usepackage{amsmath,amsfonts,amssymb}
\usepackage[ruled]{algorithm2e}
\usepackage{array}
\usepackage{textcomp}
\usepackage{stfloats}
\usepackage{url}
\usepackage{verbatim}
\usepackage{graphicx}
\usepackage{threeparttable}
\usepackage{multirow}
\usepackage{bbm} 
\usepackage{subfigure}
\usepackage{bbding}
\usepackage[T1]{fontenc}
\usepackage{color}
\usepackage{diagbox}
\usepackage{enumerate}
\usepackage{epstopdf}
\usepackage{makecell} 
\usepackage{multirow}
\usepackage{booktabs}

\begin{document}

	\begin{frontmatter}
		
		\title{Listen As You Wish: Audio based Event Detection via Text-to-Audio Grounding in Smart Cities}
		
		\author[mymainaddress]{Haoyu Tang}
		\ead{tanghao258@sdu.edu.cn}
  
		\author[mymainaddress]{Yunxiao Wang}
		
		\author[xianjiaoda]{Jihua  Zhu}
  
		\author[mymainaddress]{Shuaike Zhang}
  
		\author[mymainaddress]{Mingzhu Xu}

        \author[fuzhouda]{Qinghai Zheng}
        
        \author[mymainaddress]{Yupeng Hu \corref{mycorrespondingauthor}}
		\ead{huyupeng@sdu.edu.cn}
        \cortext[mycorrespondingauthor]{Corresponding author}
        
		\address[mymainaddress]{School of Software Engineering, Shandong University, Jinan 250101, China.}
		\address[fuzhouda]{School of Software Engineering, Fuzhou University, Fuzhou 350108, China.}
		\address[xianjiaoda]{School of Software Engineering, Xi’an Jiaotong University, Xian 710049, China.}
    \begin{abstract}
With the development of internet of things technologies, tremendous sensor audio data has been produced, which poses great challenges to audio-based event detection in smart cities. In this paper, we target a challenging audio-based event detection task, namely, text-to-audio grounding, which aims to find the exact sound segment corresponding to the target event described by a natural language query. In addition to precisely
localizing all of the desired on- and off-sets in the untrimmed audio, this challenging new task requires extensive acoustic and linguistic comprehension as well as the reasoning for the crossmodal matching relations between the audio and query. The current approaches often treat the query as an entire one through a global query representation in order to address those issues. We contend that this strategy has several drawbacks. Firstly, the interactions between the query and the audio are not fully utilized. Secondly, it has not distinguished the importance of different keywords in a query. In addition, since the audio clips are of arbitrary lengths, there exist many segments which are irrelevant to the query but have not been filtered out in the approach. This further hinders the effective grounding of desired segments.

Motivated by the above concerns, a novel Cross-modal Graph Interaction (CGI) model is proposed to comprehensively model the relations between the words in a query through a novel language graph. To capture the fine-grained relevances between the audio and query, a cross-modal attention module is introduced to generate snippet-specific query representations and automatically assign higher weights to keywords with more important semantics. Furthermore, we develop a cross-gating module for the audio and query to weaken irrelevant parts and emphasize the important ones. On the public Audiogrounding benchmark dataset, we extensively evaluate the proposed CGI model with significant improvements over several state-of-theart methods. The ablation studies demonstrate the consistent effectiveness of different modules in our model.
		\end{abstract}
		
		
	\end{frontmatter}

\section{Introduction}



\begin{figure*}[htbp]
	\centering
	\includegraphics[width=0.95\textwidth]{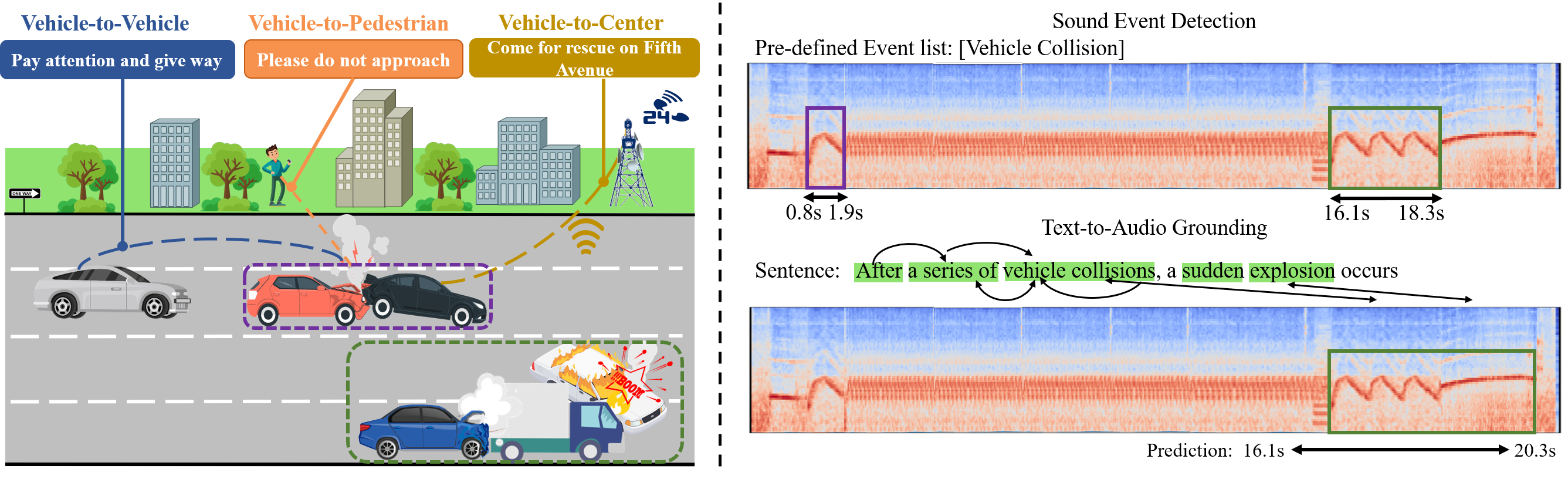}
	\caption{Illustration of the traffic event detection in VRCS. In this example, the left portion reveal that the vehicle-to-vehicle, vehicle-to-pedestrian, and vehicle-to-center technologies can be employed to ensure the safety control of vehicles and collaborative management of roadways after the vehicle collisions happens. The right portion presents the difference between SED and TAG in VRCS. The SED method can localize two ``vehicle collision'' segments in audio, yet it fails to differentiate between a singular `` vehicle collision'' and ``a series of vehicle collisions''. In contrast, given the query ``After a series of vehicle collisions, a sudden explosion occurs.'', the TAG precisely grounds the audio segment containing the events ``a series of vehicle collisions'' and ``explosion'' with the on- and off-set (i.e., from 16.1s to 20.3s).}

	\label{fig:introduction}
\end{figure*}

Nowadays, advances in the Internet of Things (IoT) technologies have driven the growth of smart devices that have significantly changed daily life in smart cities~\cite{ding2019survey,qi2020overview,ma2019audio,passos2023multimodal,aslam2022detecting}. Vehicle Road Cooperation System (VRCS), as a crucial component of smart cities, endeavors to achieve effective coordination of people, vehicles, and roads through data communication and computation. Namely, it enables proactive safety control of vehicles and collaborative management of roadways, ultimately leading to the establishment of a safe and orderly road traffic environment~\cite{You2022Safety}. Existing VRCS research ~\cite{Chen2023Cooperative} points out accurate and timely traffic event detection is a prerequisite to improve overall vehicle road cooperative control. Taking the traffic event detection in Fig.~\ref{fig:introduction} as an example, when vehicle collision accident (marked with purple box) is detected, we can not only utilize vehicle-to-vehicle and vehicle-to-pedestrian technology to alert nearby vehicles and pedestrians to take evasive actions but also vehicle-to-center technology to report the traffic accident to the traffic management center and call for road rescue assistance. Therefore, how to establish an effective recognition of traffic events is essential to achieve system-level vehicle road cooperation\cite{stiller2011information}. 

Considering the massive and heterogeneous (video, images, audio, etc.) sensor data present in modern VRCS~\cite{Lu2014Connected}, selecting appropriate data features and efficiently accomplishing traffic event detection is a nontrivial task~\cite{Sun2021Cloud}. Existing research indicates that visual sensing data (such as videos and images) is susceptible to degradation due to lighting conditions and shooting angles~\cite{chandrakala2019environmental},  leading to a deterioration in the performance of vision-based traffic event detection. Therefore, researchers are making efforts to utilize audio sensor data to accomplish traffic event detection. Audio signals emitted by vehicles and roads can provide valuable information, such as warning honks or approaching vehicle sounds~\cite{Xia2019Improving, Wu2022Environmental}. Early research focuses on the sound event detection (SED) to find and classify the special traffic events in a given audio \cite{wu2019audio,wu2020audio}. As shown in Fig.~\ref{fig:introduction}, with the predefined sound action behavior "vehicle collision", two traffic accidents (marked with purple and green boxes) can be detected. However, these SED methods cannot differentiate between the two traffic accidents, which hinders the effectiveness of subsequent rescue operations. More seriously, they are limited to the predefined sound action list, failing to identify complex activities in real traffic scenarios. 

Recently, the task of text-to-audio grounding (TAG) \cite{xu2021text} has been proposed to overcome the limitation of SED. Particularly, given a language sentence as the query, TAG aims at localizing all the audio segments in an untrimmed audio corresponding to the sound event mentioned in the query. Compared to SED, TAG is much more challenging in the traffic scenario since the queries can be arbitrary complex language descriptions, which are always sophisticated and complex. Figure \ref{fig:introduction} shows a query with its corresponding segments in the audio captured from a car crash scenario. Taking the query "After a series of vehicle collisions, a sudden explosion occurs" as an example, this typical language query emphasizes that an event of `` `vehicle collision' continuously happens shortly, followed by an `explosion' '' occurs in the audio. To successfully localize this query in audio, the model returning only `vehicle collision' is not satisfactory. Particularly, grounding such query needs to not only retrieval the segment with ``vehicle collision'' event happening in a series shortly, but also ensure that the sound of an explosion is exactly followed in the segment. To achieve this goal, the following factors are crucial: 1) Well comprehending the sophisticated query semantics by attending to the most useful word and modeling the local and non-local relations between words in the query; 2) Understanding the audio contents by weakening the expressiveness of the irrelevant parts ({e.g., occurs}) in it; 3) Aligning the audio content and query semantics by capturing their fine-grained interactions.

We have to note that the existing method simply processes the whole language query into a word encoder to construct one feature embedding as the global representation of the query \cite{xu2021text}. Despite its great performance, simply encoding the query holistically as one global feature may overlook the implicit relations between words and the keywords that provide rich semantics. In other words, this method fails to find the relationship between the words ``a series of'' and ``vehicle collisions'' as in Fig.\ref{fig:introduction}. These factors are critical to localize the audio segments containing sequential  vehicle collisions within a brief time, since the audio may contain segments of an individual vehicle collision. In addition, this method directly matches the global query feature and the audio snippets features, which cannot bridge the fine-grained matching relevance between the audio snippets and the query words.
As we can see, despite crucial for precisely grounding the desired moment, these aforementioned factors have been largely untapped in the existing method.

Considering those factors, we propose a novel Cross-modal Graph Interaction (CGI) model. In specific, after separately encoding the audio and query input into the snippet- and word-level representations, we construct a novel intra-modal query graph that regards each word as a node and explores their local and non-local implicit relations. To fully explore the cross-modal relevance in the fine-grained level, a multimodal attention mechanism is employed to enrich the matching information between these two modalities. Besides, we present a cross-gating module that automatically assigns different importance weights to audio snippets and words depending on their relevance. The key contributions of this work are four-fold:
\begin{itemize}
	\vspace{-2pt}
	\item We emphasize the importance of TAG based traffic event detection for intelligent vehicle road cooperation and highlight three crucial challenges of TAG: (1) the implicit relations of words in query; (2) weakening the expressiveness of the irrelevant parts in audio; (3) the fine-grained interactions between audio snippets and words in the query.
	\item We introduce an intra-modal query graph network to model the local and non-local relations between words in the query, and incorporate an attention module to exploit the fine-grained interactions between two modalities.
	\item We present a cross-gating scheme to emphasize the critical audio and query cues and further weaken the inessential ones based on their relevance to each other.
	\item We conduct extensive experiments on AudioGrounding dataset \cite{xu2021text} to verify the superiority of our proposed CGI model over several state-of-the-art baselines.
\end{itemize}

\section{Related Work}
In this section, we provide a concise overview of three closely related research directions as follows.

\subsection{Sound Event Detection in Smart Cities}
Audio understanding of IoT is an emerging research topic, where sound event detection (SED) in smart cities has great potential for many IoT applications \cite{imoto2013user,gerstoft2021audio}. The objective of this task is to identify the on-set and off-set of events in the audio, and also provide their event class from a pre-defined set. The early methods in this field directly detect the sound events via fully connected layers \cite{cakir2015polyphonic}. With the prevailing deep learning paradigm in the computer vision field, some researchers adopted the convolutional neural network architecture to obtain the suitable temporal frequency representations of audio \cite{zhang2015robust,salamon2017deep}. Due to the sequential nature of audio, recurrent neural networks are also employed to learn the long-term feature representation of audio \cite{parascandolo2016recurrent}. By integrating both CNN and RNN layers, some methods proposed the convolutional recurrent neural networks (CRNNs) for better SED performance \cite{xu2017convolutional}. Besides, due to the success of Transformer in many fields, Miyazaki et al. designed a Transformer-based network to encode the audio in a weakly supervised fashion \cite{miyazaki2020weakly}. More recently, some researchers introduced the task of estimating the direction-of-arrival (DOA) to SED \cite{nguyen2020sequence}, while others also proposed a new evaluation metric \cite{bilen2020framework}, both of which make this task more challenging.

Although these methods have achieved great progress for the SED task, they are still restricted to a pre-defined event set. Recently, Xu et al. \cite{xu2021text} proposed to use language queries to localize events in the audio, i.e., the text-to-audio grounding (TAG). In their model, the entire sentence query is encoded as a global feature by average-pooling, which cannot comprehensively obtain the semantics of the sentence. Besides, the common evaluation metrics of SED are adopted to verify their TAG model since both tasks require localization in the audio. Different from their method, the proposed CGI network models the implicit relations in the query and further captures the fine-grained interactions between cross-modal features, which thus achieves better localization performances.

\subsection{Language Grounding in Visual Data}
There are some language grounding tasks in computer vision (CV) field that focus on localizing language sentences in visual data. Two mainstream grounding tasks are included here, namely, image grounding \cite{rohrbach2016grounding} and temporal language grounding \cite{Gao_2017_ICCV}. Given a sentence query, their target is to localize an image region or a video moment in an image or video, respectively. Obviously, modeling pairwise relations between words in queries and capturing cross-modal interactions are also important for those tasks. For example, Chen et al. \cite{chen2018temporally} proposed a Match-LSTM structure to match the sentence and video for the temporal language grounding task; Liu et al. \cite{liu2018cross} employed the query attention module to adaptively reweight the features of each word in query according to the video content. For the image grounding task, Mu et al.~\cite{mu2021disentangled} built a scene graph to capture different motifs in the image and then devised a disentangled graph network, which integrates the motif contextual information into image representations.

Different from them, our proposed CGI model addresses the TAG task, which captures the interactions between sentence query and audio instead of visual data. In addition, for a given query, while the language sentence in visual data only needs to ground one object (e.g., an image region or a video moment), our method often needs to return more than one corresponding segment in a single audio, which is much more challenging. Moreover, apart from the query graph and attention module for query modeling that have been studied in those methods \cite{liu2020jointly,tang2021frame}, we also present a cross-gating mechanism, which can highlight the critical parts in audios and queries, which can further enhance their representations. 

\subsection{Graph Neural Networks in Language}

Extended from the random walk based methods, graph neural network (GNN) \cite{scarselli2008graph} has drawn much research attention recently. It is often used to process the sequential information of graph-structured data in recommendation systems. Due to the graph-structured property of natural language, GNN is also adopted to exploit the semantic relations of language sentence \cite{beck2018graph,marcheggiani2017encoding}. As shown in many methods, the semantic information can be successfully captured when GNN is incorporated for language modeling \cite{zhang-etal-2018-graph,huang2020aligned}. Considering the great progress GNN has made, we introduce a novel intra-modal query graph to propagate the semantic messages in our model, which captures the high-order relations in query and further enriches the query features for precise text-to-audio grounding.

\begin{figure*}[htbp]
	\centering
	\includegraphics[width=0.95\textwidth]{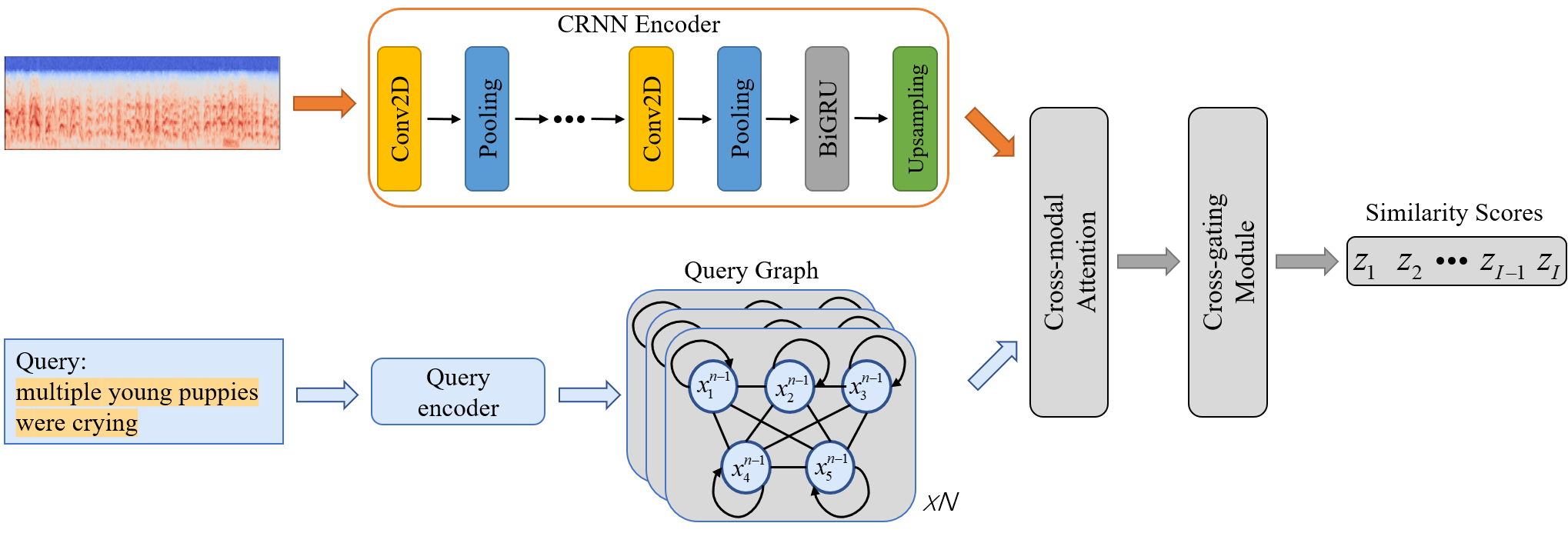}
	\caption{The structure of the proposed CGI model, which consists of the following components: two encoders that extract the audio and query features separately; an intra-modal query graph to model the implicit information between words in query; a cross-modal attention module for the fine-grained semantics; a cross-gating module which automatically emphasizes crucial parts in audio and query; a grounding module to measure the similarities between the audio snippets and query features for precisely returning the desired audio segments.}
	\label{fig:network}
\end{figure*}

\section{The Proposed CGI Model}
\subsection{Problem Formulation}
We formulate the text-to-audio grounding task and then demonstrate the detail of the CGI model as in Figure \ref{fig:network}. We denote an untrimmed audio as $U=\{{{u}_{i}}\}_{i=1}^{I}$, where ${u}_{i}$ is $i$-th audio snippet with $I$ denotes its length. The given query is also represented word-by-word as  $S= \{ {{s}_{l}} \}_{l=1}^{L}$ with $L$ denotes the number of words in $S$. For each query-audio pair $\{U,S\}$, our model targets at grounding all the ground-truth audio segment $\{ {{t}_{s}^{c},{t}_{e}^{c}} \}_{c=1}^{C}$ in $U$, where $C$ represents the number of the ground-truth audio segments in $U$. ${t}_{s}^{c}$ and ${t}_{e}^{c}$ denotes the on-set and off-set for $c$-th ground-truth audio segment.

\subsection{Encoder}
To encode each input audio sequence $U$, the standard Log Mel Spectrogram (LMS) audio feature extractor is adopted as:
\begin{equation}
	\label{equation:audioencode}
	\mathbf{U}_{a} = \operatorname{LMS}(U) 
\end{equation}
where $\mathbf{U}_{a} \in \mathbb{R}^{L \times m}$ is the encoded audio embeddings with $L$ denoting its length and  $m$ as its dimension. A convolutional recurrent neural network (CRNN) \cite{xu2020crnn} is then employed to encoder the audio feature $\mathbf{U}_{a}$, which contains five padded $3 \times 3$ convolution blocks. After a followed bidirectional gated recurrent unit (BiGRU) that processes more sequential contextual information in the audio, an upsampling operation is adopted to restore the temporal dimension to the same length $I$ as the original audio feature. Namely, $ \mathbf{U}=\{\boldsymbol{u}_{i}\}_{i=1}^{I} \in \mathbb{R}^{I \times d}$ denotes the output feature of the CRNN encoder, with $d$ denoting each feature embedding dimension.

As for the sentence query, a look-up vocabulary is adopted to transfer each word ${s}_{l}$ in the query to a word embedding $\mathbf{s}_{l}$, resulting in the query feature $\mathbf{S}= \{ {\mathbf{s}_{l}} \}_{l=1}^{L} \in \mathbb{R}^{L \times d}$.
\subsection{Intra-modal Query Graph}
Since the grounding of audio segments needs to understand the query description, it is necessary to fully model the intra-modal relations in the sentence query. To this end, we propose an intra-modal query graph network to capture such relations between words for better query representations. Specifically, a directed sentence query graph $\mathcal{G}=(\mathcal{V}, \mathcal{E})$ is constructed. $\mathcal{V}$ contains all the words in the query as nodes, and $\mathcal{E}$ denotes the edge set containing all node pairs between words, i.e., the edge $\boldsymbol{e}_{(l, j)}$ denotes the relation from the $l$-th word node to the $j$-th word node, and $\boldsymbol{e}_{(j, l)}$ denotes the reverse relation. With $N$ layers of such graph stacked together,  the comprehensive intra-modal relations in the query can be captured. In the next, we will describe the message aggregation and updating process of the $n$-th query graph layer.

\subsubsection{Message Passing and Aggregation}
We first adopt the output of the last query encoder $\mathbf{S}= \{ {\mathbf{s}_{l}} \}_{l=1}^{L}$ as the initialized representations for each word node in the first graph layer, referred to as $ \boldsymbol{X}^{0}=  \{ {\mathbf{s}_{l}} \}_{l=1}^{L}$. Accordingly, the node features of ($n-1$)-th layer is denoted as $\boldsymbol{X}^{n-1}= \{ \boldsymbol{x}_{l}^{n-1} \}_{l=1}^{L}$.  Considering that for a word node, the neighboring words are usually more important than the distant ones in the query. Therefore, the position information of each word node in query sequence should also be integrated into the node representations to better capture the node interactions. To provide such position notions, the positional encodings (PE) is appended to the corresponding node features, which has been widely used for the language representations like Transformer \cite{vaswani2017attention} in some language processing tasks. Formally, for the $l$-th node feature $\boldsymbol{x}_{l}$ of the query, the positional encoding is defined as:
\begin{equation}
	\operatorname{PE}\left(\boldsymbol{x}_{l}\right)=\left\{\begin{array}{ll}
		\sin \left(l / M^{k / d}\right), & \text { if } k \text { is even } \\
		\cos \left(l / M^{k / d}\right), & \text{ otherwise}
	\end{array}\right.
\end{equation}
where  $k$ is the feature index varing from 1 to $d$. $M$ is a  scalar constant ,which is  empirically set to 10000. After the node features and position encodings are integrated, the edge weights between the paired nodes $\boldsymbol{x}_{l}^{n-1}$ and $\boldsymbol{x}_{j}^{n-1}$ can be calculated as:
\begin{equation}
	{{a}}_{lj}^{n} =\left(\boldsymbol{x}_{l}^{n-1}+\lambda_{1} \operatorname{PE}\left(\boldsymbol{x}_{l}\right)\right)\left(\boldsymbol{x}_{j}^{n-1}+\lambda_{1} \mathrm{PE}(\boldsymbol{x}_{j})\right)^{T}
\end{equation}
where $\boldsymbol{x}_{l}^{n-1}$ represents the $l$-th node feature at ($n-1$)-th graph layer, and $\lambda_{1}$ balances the contribution of the position information. 

To aggregate the messages passed from word nodes,  We integrate the features of all word nodes for each word node in the edge-weighted manner. Specifically, this message aggregation is formulated as follows:
\begin{equation}
	\alpha^{n}_{lj}=\frac{\exp \left({{a}}_{lj}^{n}\right)}{\sum_{j=1}^{L} \exp \left({{a}}_{lj}^{n}\right)}
\end{equation}
\begin{equation}
	\boldsymbol{h}^{n}_{l}=\sum_{j=1}^{L}\alpha^{n}_{lj}\boldsymbol{x}_{j}^{n-1}\in \mathbb{R}^{1 \times d}
\end{equation}
where  $\boldsymbol{h}^{n}_{l}$ is the $l$-th node's aggregated message for subsequent updating, and $\alpha^{n}_{lj}$ is the normalized weight between $\boldsymbol{x}_{l}^{n-1}$ and $\boldsymbol{x}_{j}^{n-1}$ by a softmax function. 

\subsubsection{Update of node representations} After the process of aggregating the message from all its neighbors, the new node representation at $n$-th graph layer is obtained by considering its feature at the prior layer and the received messages. More formally, this updating processing can be expressed as:
\begin{equation}
	\boldsymbol{x}_{l}^{n}=\operatorname{F}\left(\boldsymbol{x}_{l}^{n-1}, \boldsymbol{h}^{n}_{l}\right)
\end{equation}
where $\operatorname{F}$ is an updating function which fuses the prior node feature and the received messages. Usually, there are two common forms of this update equation: 1) The element-wise matrix addition on $ \boldsymbol{x}_{l}^{n-1}$ and $\boldsymbol{h}^{n}_{l}$ which directly incorporates the previous node and the aggregated messages; 2) The concatnation of $ \boldsymbol{x}_{l}^{n-1}$ and $\boldsymbol{h}^{n}_{l}$ which mainly focuses on retaining their own information. Instead of these two operations, we adopt a ConvGRU layer to update the node feature as in \cite{liu2020jointly}. As a convolutional counterpart to original GRU, such ConvGRU layer can perserve the sequential information of $ \boldsymbol{x}_{l}^{n-1}$ and $\boldsymbol{h}^{n}_{l}$. After the updating process, all the updated word nodes  $\boldsymbol{X}^{n}= \{ \boldsymbol{x}_{l}^{n} \}_{l=1}^{L}$ are fed into the next query graph layer for further message passing. Finally, the last output of the $N$-th layer  is referred to as $\boldsymbol{X}^{N} \in \mathbb{R}^{L \times d}$, which will be used for cross-modal interaction.

\subsection{Cross-Modal Interaction}
\subsubsection{Attention Module}
With the intra-modal graph module, the enriched word representations $\boldsymbol{X}^{N}$ which fully explore the semantic information in the query are obtained. In the next, we need to learn fine-grained query representations through cross-modal interaction. As in Xu et al. \cite{xu2021text}, the direct method is to average all the embeddings from the words in the query. Such operation treats these words equally for the global query representation. However, it is noted that based on the uniqueness of the required audio segment, the importance of the words in a query differs from each other for the final localizations. For example, given a query ``an ambulance sounds the siren'', the word `ambulance' and `siren' conveys more semantics than other words, which should thus be paid more attention. As a result, it is crucial to adopt an attention module to distinguish the importance of the words in query.

To this end, we devise an attention module that explores the snippet-by-word interactions for snippet-specific query representations. Formally, the attention weights between each pair of the audio snippet and word feature are computed as:
\begin{equation}
	\begin{gathered}
		r_{li}=\mathbf{w}_{r}^{T} \cdot \operatorname{tanh} \left(\mathbf{W}_{s} \boldsymbol{x}_{l}^{N}+\mathbf{W}_{a} \boldsymbol{u}_{i}+\mathbf{b}_{r}\right) \\
	\end{gathered}
\end{equation}
where $\operatorname{tanh}$ denotes non-linear tanh function. $\mathbf{w}_{r}^{T}$, $\mathbf{W}_{s}$, $\mathbf{W}_{a}$, and $\mathbf{b}_{r}$ are the learnable parameters. The attention score $r_{li}$ describes the similarity between $i$-th audio snippet and $l$-th word. The snippet-specific query feature for $i$-th audio snippet is obtained by a weighted summarization of all the word features in the query as:
\begin{equation}
	\bar{\mathbf{s}}_{i}=\sum_{l=1}^{L}\operatorname{Softmax}_{c}(\boldsymbol{r}) \cdot  \boldsymbol{x}_{l}^{N}
\end{equation}
where $\operatorname{Softmax}_{c}$ denotes the softmax function along the column of a matrix. The obtained snippet-specific features of all snippets are concatenated together as $\bar{\mathbf{S}}= \{ \bar{\mathbf{s}}_{i} \}_{i=1}^{I} \in \mathbb{R}^{I \times d}$ for subsequent cross gating.

\subsubsection{Cross-gating Module}
\begin{figure}[htbp]
	\centering
	\includegraphics[width=0.95\columnwidth]{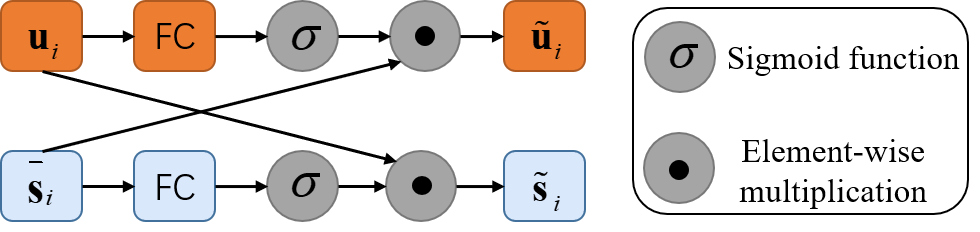}		
	\caption{The structure of the cross-gating module.}
	\label{fig:Cross_Gating}
\end{figure}

Based on the snippet-specific query representation $\bar{\mathbf{S}}$ and the audio feature $ \mathbf{U}=\{\boldsymbol{u}_{i}\}_{i=1}^{I}$, a cross-gating module \cite{feng2018video} is proposed to automatically calculate the different importance weights of both the most relevant parts and the inessential ones. Specifically, the gating of query features depends on the audio features, and the audio streams are gated by the corresponding sentence query feature. As shown in \ref{fig:Cross_Gating}, the detailed cross-gating scheme is formulated as follows:
\begin{equation}
	\begin{array}{ll}
		k_{i}^{u}=\sigma\left(W_{u}^{g} \boldsymbol{u}_{i}+b_{r}^{g}\right), & \tilde{\mathbf{s}}_{i}=\bar{\mathbf{s}}_{i} \odot k_{i}^{u} \\
		k_{i}^{s}=\sigma\left(W_{s}^{g} \bar{\mathbf{s}}_{i}+b_{q}^{g}\right), & \tilde{\boldsymbol{u}}_{i}=\boldsymbol{u}_{i} \odot k_{i}^{s}
	\end{array}
\end{equation}
where $\sigma$ denotes the sigmoid activation function and $\odot$ denotes the dot production. $W_{u}^{g}, W_{s}^{g} \in \mathbb{R}^{d \times d}$, and $b_{u}^{g}, b_{s}^{g} \in \mathbb{R}^{d \times 1}$ represent the trainable parameters. It can be observed from these functions that the cross-gating mechanism controls the degree of interactions of one modality with the other. On the one hand, if the audio feature $\boldsymbol{u}_{i}$ is unrelated to the query feature $\bar{\mathbf{s}}_{i}$, both the audio and query representation $\boldsymbol{u}_{i}$ and $\bar{\mathbf{s}}_{i}$ are gated out to reduce their influence on the subsequent grounding. On the other hand, if they are closely related, this mechanism is capable of enriching their cross-modal interactions.

\subsection{Grounding and Learning}
The grounding module of the proposed CGI model is employed in this section, which works on the obtained features of two modalities to estimate the specific on-set and off-set in the audio for TAG task. Considering that the audio snippets which have greater correlations with the corresponding query are more likely to be the desired grounding results, we directly compute the similarities between the audio snippets and sentence query features to obtain a vector with the audio length $I$ as follows:
\begin{equation}
	z_{i}=\operatorname{sim}\left(\tilde{\boldsymbol{u}}_{i},\tilde{\mathbf{s}}_{i}\right)=\exp \left(-\left\|\tilde{\boldsymbol{u}}_{i}-\tilde{\mathbf{s}}_{i}\right\|_{2}\right)
\end{equation} 
Following the previous method \cite{xu2021text}, the binary cross-entropy (BCE) loss is applied as the training criterion, which is calculated as:
\begin{equation}
	\mathcal{L}_{\mathrm{BCE}}=-\frac{1}{I} \sum_{i=1}^{I} {y}_{i} \cdot \log \left(z_{i}\right)+\left(1-y_{i}\right) \cdot \log \left(1-z_{i}\right)
\end{equation}
where $y_{i}$ is the ground-truth label for $i$-th audio embedding. When $y_i$ is either 1 or 0, it indicates whether the query is present in the $i$-th audio embedding. During the evaluation stage, the similarity vector $\boldsymbol{z}$ is binarized to a prediction vector $\boldsymbol{\hat{y}}$ through the threshold $\beta$ as:
\begin{equation}
	\hat{y}_{i}=\left\{\begin{array}{l}
		1, \text{ if } \, z_{i}>\beta \\
		0, \text { otherwise }
	\end{array}\right.
\end{equation}

\section{Experiment}
Extensive experiments on the AudioGrounding dataset have been conducted. The experimental settings for all other methods, such as the hyperparameters and the training settings, are the same as what they have reported in their papers.

\subsection{Datasets}
\textbf{AudioGrounding}: This dataset was constructed based on the Audioset \cite{7952261} and AudioCaps \cite{kim-etal-2019-audiocaps} dataset for Automated Audio Captioning task by Xu et al. \cite{xu2021text}. The AudioGrounding dataset contains 4,590 audios and 4994 sentence captions, resulting in total 13,985 audio-query pairs. Following the same split settings \cite{xu2021text}, we split these 13,985 pairs into three separate parts: 12373, 451, and 1161 for training, validating, and testing to verify the CGI model for the TAG task.

\subsection{Implementation Details}
The queries are first lowercased and then tokenized by the standard Stanford CoreNLP \cite{manning-EtAl:2014:P14-5} tool. The word embeddings are adopted to embed each word to a 256-dimension feature. For the raw audios, we extract the LMS feature of 64 dimensions, employing a window with 40ms size and 20ms shift, resulting in the audio embedding $\mathbf{U}_{a} \in \mathbb{R}^{L \times 64}$. The size of the embeddings $d$ in our model is set to 256. We train the proposed CGI model end to end with at most 100 epochs for the batch size of 64, where the early-stopping strategy is employed. The Adam optimization algorithm is adopted, and the learning rate is set to 0.001, which will be gradually decreased by 10 if the validating loss does not improve for five epochs. During the evaluation, the threshold $\beta$ is set to 0.4.

\subsection{Evaluation Metrics}
To verify our CGI model, we use several standard metrics which are commonly used for the SED task, following the previous method \cite{xu2021text}. To value the smoothness of prediction segments and penalize irrelevant predictions, the event-based metrics are adopted, including precision, recall, and F1, which are denoted as P, R, and F$_{1}$, respectively. It is noting that for the F$_{1}$ score, the t-collar value of 100 ms, and the 20\% tolerance between the ground-truth and prediction audio segments are adopted. Since the performance of these metrics is relevant to the threshold, we also compute the more robust threshold-independent polyphonic sound detection score (PSDS), and the hyperparameters of PSDS are defaulted as $\rho_{\mathrm{DTC}}=\rho_{\mathrm{GTC}}=0.5$, $\rho_{\mathrm{CTTC}}=0.3$, $\alpha_{\mathrm{CT}}=\alpha_{\mathrm{ST}}=0.0$, $e_{\max }=100$.

\subsection{Comparison with State-of-the-Arts}
As in Table \ref{table:results on Comparison with State-of-the-Arts}, our CGI model is compared on the AudioGrounding dataset with those following state-of-the-art methods where the best results of each metric are highlighted in bold:
\begin{itemize}
	\vspace{-2pt}
	\item \textbf{TGA} \cite{xu2021text}: This model is designed for text-to-audio grounding by integrating the global query feature and audio features. It outputs the similarity vector between the audio snippets features and the mean-pooled sentence query feature, and then grounds the desired segments by a threshold upon the similarity vector.  
	\item \textbf{Attention-query}: This designed baseline first separately encodes the audio and query as our CGI model, and then learns the query attention based on the audio content. Specifically, the encoded audio features  $\mathbf{U} \in \mathbb{R}^{I \times d}$ are averaged to obtain a global audio representation $\mathbf{u}_{g} \in \mathbb{R}^{d}$, which is employed to compute the similarity score with each word $\mathbf{s}_{l}$ in query $\mathbf{S} \in \mathbb{R}^{L \times d}$. After reweighting and summarizing all word features by the similarity scores, the global query feature $\mathbf{s}_{g}$ is obtained.   

	\item \textbf{Match-LSTM} \cite{wang2016learning}: This method introduced a Match-LSTM structure that learns the fine-grained interactions between question and document for the machine comprehension task. Considering the similarity between this task and TAG, we design a baseline by regarding the query and the audio as the question and document, respectively, and the Match-LSTM structure is directly employed to capture the matching relations between the obtained audio and sentence query features. 
\end{itemize} 

It is noting that the same grounding and learning module is adopted as ours for the Attention-query and Match-LSTM baseline.

\begin{table}
	\caption{Performance comparison on AudioGrounding dataset.}
	\label{table:results on Comparison with State-of-the-Arts}
	\centering
	\begin{tabular}{c||c|c|c|c}
		\hline
		Method & P & R& F$_{1}$& PSDS\\
		\hline
		Random & 0.02&1.56&0.04&0.00\\
		TAG  & 28.60&27.90&28.30&14.70\\
		Attention-query  & 27.35&31.00&29.06&18.61\\
		Match-LSTM  & 30.25&34.65&32.30&20.03\\
		CGI& \textbf{31.27}&\textbf{36.57}&\textbf{33.72}&\textbf{22.81}\\
		\hline
	\end{tabular}
\end{table}

From the results, the following observations stand out. First, although inferior to the TAG baseline in Precision, the designed baseline Attention-query already performs much better than the TAG baseline in all other metrics, which strongly verifies the effectiveness of attending to the keywords in the query. The designed Match-LSTM method achieves even better performance over Attention-query method in all metrics because it not only attends the useful words in the query but also captures the relations between the audio snippets and query words. Compared to those methods, our CGI model achieves the best performance on AudioGrounding dataset. Specifically, it consistently surpasses all the baselines by a large margin in all evaluation metrics. Although the precision and recall metrics are contradictory to some extent, the CGI model still achieves around 1.0\% and 2.0\% absolute improvements over the Match-LSTM method. For the PSDS metric which is independent of the threshold $\beta$, the CGI model brings a 2.8\% absolute improvement compared to the second best Match-LSTM method. Overall, the excellent results of the CGI model are obtained by the combining effects of the intra-model graph for query modeling, the cross-modal attention for fine-grained interactions, and the employment of the cross-gating scheme for representations enrichment. 

\begin{table*}
	\caption{Evaluation results of ablation study for the proposed CGI model on the AudioGrounding datasets where QG, CMT, CG, PE denote the query graph module, the cross-modal attention module, the cross-gating scheme, and the positional encoding, respectively. In this table, the ``\checkmark" symbol indicates that the variant model enables the corresponding component.}
	\label{table:ablation on anetcap}
	\centering
	\scalebox{0.95}{\begin{tabular}{c|c|c|c|c|c|c|c|c}
		\hline
		Method&CMT&QG&PE&CG& P & R& F$_{1}$& PSDS\\
		\hline\hline
		TAG&&&&&28.60&27.90&28.30&14.70\\
		CGI(w/o.QG and CG)&\checkmark&&&&29.45&31.83&30.60&18.63\\
		CGI(w/o.QG)&\checkmark&&&\checkmark&30.79&34.53&32.55&21.47\\
		CGI(w/o.PE)&\checkmark&\checkmark&&\checkmark&31.04&34.85&32.83&22.33\\
		CGI(w/o.CG)&\checkmark&\checkmark&\checkmark&&\textbf{31.78}&34.11&32.90&22.36\\ 
		CGI(Full)&\checkmark&\checkmark&\checkmark&\checkmark&31.27&\textbf{36.57}&\textbf{33.72}&\textbf{22.81}\\
		\hline
	\end{tabular}}
\end{table*}

\subsection{Ablation Study}
We investigate the contribution of all the components in our CGI model by the ablation studies, including the intra-modal query graph, the positional encoding, the cross-modal attention module, and the cross-gating module. Specifically, the following variants of our model are generated by removing one or two components at a time. TAG is used as the baseline here.
\begin{itemize}
	\vspace{-2pt}
	
	\item CGI (w/o.CG): We eliminate the cross-gating module of our model. That is, the outputs of the cross-modal attention module are directly adopted for learning and grounding.
	\item CGI (w/o.PE): For the full model, we remove from the query graph the positional encoding parts that provide temporal sequential information to nodes. 
	\item CGI (w/o.QG): We then disgard the intra-modal query graph from the full CGI model. Note that the PE is naturally removed since it is employed in the query graph.
	\item CGI (w/o.QG and CG): We finally remove the query graph and cross-gating together, and only use the cross-modal attention module.  
\end{itemize}

We compare these variants of our model on the AudioGrounding task, and the ablation results are shown in Table \ref{table:ablation on anetcap}, where the best results are highlighted in bold and the enabled components in the model variants are marked with a ``$\checkmark$'' symbol. From these ablation results, the following conclusions stand out:
\begin{itemize}
	\vspace{-2pt} 
	\item  First, CGI (w/o.QG and CG) shows consistent improvements over the TAG baseline, indicating that capturing the matching relations between audio snippets and words in the sentence query is beneficial to strengthen the cross-modal alignment and further enrich the expressiveness of the model. 
	\item  Jointly analyzing the results of CGI (w/o.PE) and CGI (w/o.QG) variants, we find that discarding the query graph overlooks the implicit relations between words which is crucial for the query comprehension and modeling, and thus degrades the grounding performance, especially in term of PSDS metric. 
	\item The proposed CGI model outperforms both CGI (w/o.CG) and CGI (w/o.PE) variant models in nearly all metrics except for the Precision metric, where the CGI model also achieves a competing result. This fact demonstrates that removing the cross-gating module hurts the representations of meaningful parts in query and audio, and ignoring the positional encoding will lose temporal contextual information in the query and further hurts the performance.
	\item Finally, almost all the variants of our model yield better performance than all compared methods in Table \ref{table:results on Comparison with State-of-the-Arts}, which verifies that the great performance of our model does not depend on a single component but their combining effects.
\end{itemize}

\begin{figure}[htbp]
	\centering
	\includegraphics[width=0.90\columnwidth]{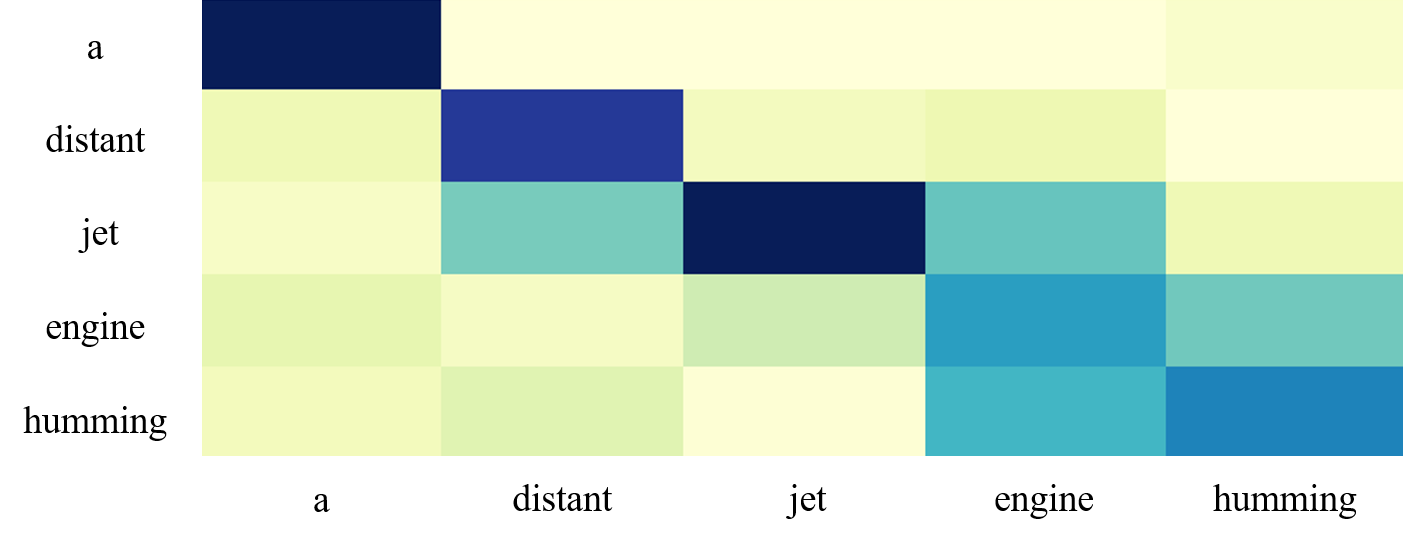}		
	\caption{Visualization of the edge weights after the softmax function along the column in the intra-modal query graph, where the dark the dot color is, the larger the related edge weight value is.}
	\label{fig:visualization_graph}
\end{figure}

\subsection{Qualitative Results} 
In this section, we first give a visualization from the AudioGrounding dataset on the implicit word relations in the intra-modal query graph, and then show several examples of the proposed CGI method and some baselines for the TAG task.

A heatmap generated from the qualitative edge weights in the query graph of an example is illustrated in Figure \ref{fig:visualization_graph}, where the darker the dot color is, the larger the corresponding edge weight is. Note that the edge weights in this figure are obtained after feeding the edge weight matrix into the softmax equation along the column, i.e, when comparing the degree of the relation of two words with a particular word, we should compare the spots along the column of this particular word. As in this figure, for the word ``a'' which has few relations with other words, a very sharp distribution is obtained for the word ``a'', where its edge weight value is concentrated in ``a'' itself. Besides, our query graph assigns much smaller edge weights to the word ``a'' for all the other words. These facts are reasonable since this word has few relations with other words in the query. As for the word ``engine'', its edge weights are inclined to be more evenly distributed, because the word ``jet'' provides ``engine'' both the object and semantic information, and the ``humming'' sound usually also has an implicit relationship with the engine. The large edge weight of the word ``engine'' for the word ``humming" accordingly confirms the relation between them. As we can see, the query graph can assign higher weights to bridge the implicit correspondence between the words in query, which enriches the query feature and further benefits the grounding process.

\section{Conclusion}
In this paper, we highlight the importance of TAG towards intelligent vehicle road cooperation, and consider its three factors, including (1) the comprehensive semantic relations between words in the given query; (2) strengthening the crucial snippets and weakening the inessential parts in audio; (3) capturing the cross-modal interactions between words and audio snippets, and we presented a novel query graph with Cross-gating Attention (CGI) model for this task. Specifically, different from previous methods simply matching the audio snippet features with a mean-pooled query feature,  we introduce an intra-modal query graph to comprehend the relations between words, and adopt an attention module that generates the snippet-specific query representations to model the cross-modal interactions between words and audio snippets. Moreover, we present a cross-gating module that weakens the unimportant parts in audio and query to further enhance their representations. Comprehensive experimental results on the AudioGrounding dataset have verified our CGI model, where our CGI model outperforms the existing method and several designed baselines by a large margin.

In the future, we will further explore the effectiveness of TAG from the following aspects: 1) we will incorporate the Transformer architecture for modeling cross-modal information since it has been proven to be powerful to process the sequential data \cite{vaswani2017attention}; 2) we plan to introduce other graph networks for audio modeling, which will enhance the audio perception capability and thus boosting the grounding performance; and 3) we will design the lightweight TAG model to achieve the collaboration on both the edge and cloud devices for diverse vehicle road cooperatio applications.

\bibliographystyle{IEEEtran}

\vfill

\end{document}